\documentclass{PoS}

\title{Lattice calculations of $D_s$ to $\eta$ and $\eta'$ decay form factors}

\ShortTitle{Lattice calculations of $D_s$ to $\eta$ and $\eta'$ decay form factors}

\author{\speaker{Issaku Kanamori}%
	\\
       Institut f\"ur Theoretische Physik, University of Regensburg,
       D-93040 Regensburg, Germany\\
       E-mail: \email{issaku.kanamori@physik.uni-regensburg.de}}

\abstract{
We report on lattice results of the form factors 
for semi-leptonic decays of the $D_s$ meson to $\eta$ and $\eta'$, 
with $n_f=2+1$ configurations.
The calculation contains 
disconnected fermion loop diagrams, which are challenging to calculate 
on the lattice.
Our result shows that the disconnected diagrams give 
significant contributions to the form factors.
}

\FullConference{Xth Quark Confinement and the Hadron Spectrum\\
         8-12 October 2012\\
         TUM Campus Garching, Munich, Germany}

\usepackage{amsmath}

\newcommand{\DiagTwoPtConn}[2]{%
\raisebox{-0.3em}{%
\makebox[0pt]{\raisebox{1.3em}{\hspace*{4em}${#1}$}}%
\makebox[0pt]{\raisebox{-0.8em}{\hspace*{4em}${#2}$}}%
\includegraphics[width=5em]{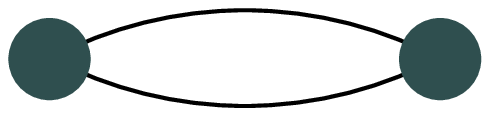}
}}

\newcommand{\DiagTwoPtDisconn}[2]{%
\raisebox{-0.2em}{%
\makebox[0pt]{\raisebox{1em}{\hspace*{3em}${#1}$}}%
\makebox[0pt]{\raisebox{1em}{\hspace*{7em}${#2}$}}%
\includegraphics[width=5em]{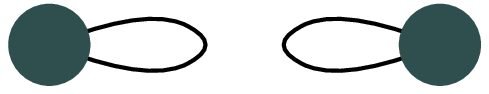}
}}

\begin{document}

\section{Introduction}

About 15\% of the decays of the $D_s$ meson are semi-leptonic.
There are by far dominated by decays with $\eta$, $\eta'$ and $\phi$
mesons in the final state.
The decays into $\eta$ and $\eta'$ contain
interesting flavor physics,
in particular,
the mixing of $\eta$ and $\eta'$.
Investigating the corresponding form factors helps to understand
the magnitude of the gluonic contribution to $\eta'$
(see \cite{DiDonato:2011kr}, for example).
The mixing can be studied through 
both $D$  and $D_s$ semi-leptonic decays.
However, the latter is easier to access in experiment,
since 
it is a Cabibbo allowed process.

The $D_s \to \eta^{(')}$ semi-leptonic decays are characterized 
by the following two
form factors, $f_0(q^2)$ and $f_+(q^2)$:
\begin{equation}
 \langle \eta^{(')}(k)| V^\mu(q^2) |D_s(p)\rangle
 = f_+(q^2) \left[ (p+k)^\mu - \frac{M_{D_s}^2 - M_{\eta^{(')}}^2}{q^2}q^\mu\right]
   + {f_0(q^2)}  \frac{M_{D_s}^2 - M_{\eta^{(')}}^2}{q^2}q^\mu,
 \label{eq:vector-current}
\end{equation}
where $V^\mu$ is a vector current and $M_{D_s}$ and $M_{\eta^{(')}}$ are
the masses of $D_s$ and $\eta^{(')}$, respectively.
Together with the CKM matrix, one can relate the form factors to the
decay width and thus the branching ratio.  
Experimental results for these decay modes are starting to appear,
see \cite{Yelton:2009aa,Yelton:2010js}. 
The matrix element (l.h.s) is what we can obtain from the lattice.
We will focus only on the scalar form factor $f_0(q^2)$, which 
can be related to the scalar matrix element \cite{Na:2009au}
\begin{equation}
 f_0(q^2) 
= \frac{m_c - m_s}{M_{D_s}^2 - M_{\eta^{(')}}^2}\langle \eta^{(')} |S| D_s \rangle,
 \label{eq:scalar-current}
\end{equation}
where $S=\bar{s} c$, and 
$m_s$ and $m_c$ are the $s$- and $c$-quark masses, respectively.
This relation has an advantage over eq.~(\ref{eq:vector-current}), in
the sense that it does not require the current renormalization.

Currently, only a prediction of the form factors from the light cone 
sum rules is available \cite{Azizi:2010zj}.
A first principles calculation on the lattice
is desirable.
It also provides an interesting playground for quantum field theory,
since physics relevant to $\eta'$ should contain non-trivial effects
from the chiral anomaly as characterized by the Veneziano-Witten formula
\cite{Witten:1979vv,Veneziano:1979ec}.

However, a lattice calculation of the decay form factor
is technically challenging, 
as it requires the evaluation of disconnected fermion loop diagrams.
In order to obtain the relevant matrix elements,
we need to compute the following three point functions shown pictorially:
\begin{equation}
\langle \eta^{(')}(\vec{k}) | S(\vec{q}) |D_s(\vec{p}) \rangle
=
\raisebox{-1em}{\includegraphics[width=8em]{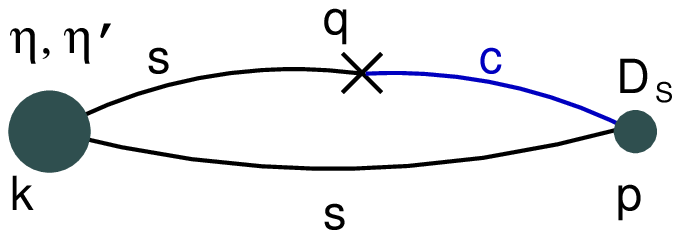}}
-\sum_{l=u,d,s}\left(
\raisebox{-1em}{\includegraphics[width=8em]{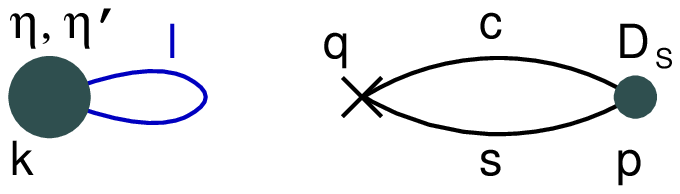}}
\right),
\label{eq:3pt}
\end{equation}
where the solid lines represent fermion propagators.
The first and the second terms are
the connected and disconnected fermion loop diagrams, respectively.
Note that the disconnected one is in fact connected by gluons.
We expect to see the effect of the anomaly 
in the pseudoscalar flavor singlet sector
and this is only possible with the disconnected terms.

To obtain a fermion propagator, we need to invert the Dirac operator,
a matrix of typically $O(10^7) \times O(10^7)$.
Many of the quark propagators in the three point function 
appearing in eq.~(\ref{eq:3pt}) only require 
a few columns of the inverted matrices.
However, the disconnected diagram requires the whole inverted matrix
(i.e. an all-to-all propagator) for the light quarks ($l$)
and thus is computationally much more expensive to calculate. 
Note that the disconnected contributions 
are summed over the three light flavors, which enhances the contributions
by roughly a factor of three.
This is another reason to expect that they may be large.

Although the calculation is challenging, 
it is still feasible \cite{Bali:2011yx}.
We use stochastic estimations of all-to-all propagators in the
disconnected fermion diagrams.
In addition, we also use the stochastic method
for the charm quark propagator in the connected 
diagram (denoted by a
blue line in eq.~(\ref{eq:3pt})) \cite{Evans:2010tg}.
This allows us to access many different
momentum combinations, which helps to improve the signal by averaging over
rotationally equivalent momenta.
The stochastic estimations introduce stochastic noise in addition to
gauge noise, and we combine several noise reduction techniques,
such as the standard 
low mode averaging and the truncated solver method (TSM)~\cite{Bali:2009hu}.

To carry out the lattice calculation, we use
QCDSF $n_f=2+1$ configurations~%
\cite{Bietenholz:2010jr, Bietenholz:2011qq}.
The strategy of these sets of configurations is to keep the
flavor singlet mass average of the three quarks,
 $\frac{1}{3}(m_u+m_d+m_s)$, constant.
Starting from the flavor SU(3) symmetric point, $m_{u,d}=m_s$, 
$m_{u,d}$ is reduced and $m_s$ is increased.
This setup is ideal to study flavor physics in the SU(3) flavor
basis.
So far, we have analyzed 939 configurations at the flavor symmetric
point, corresponding to pion and Kaon masses $M_\pi=M_K=450\, {\rm MeV}$,
and 239 configurations with $M_\pi=348\, {\rm MeV}$ and $M_K=483\,{\rm MeV}$.
The physical spacial extent of the lattice is roughly $1.9\, {\rm fm}$,
and the lattice spacing is $a\sim 0.08 {\rm\, fm}$ for 
both sets of configurations.
The stout link non-perturbatively improved clover action (SLiNC) 
\cite{Cundy:2009yy}, 
for which the discretization effects are removed to $O(a)$,
is used for both the dynamical $u,\ d$ and $s$-quarks 
and the partially quenched $c$-quark,

In the next section, we extract the masses of $\eta$ and $\eta'$,
 and a mixing angle between the
SU(3) flavor singlet-octet basis and the physical states.
We then present our result for the scalar form factor $f_0$ 
in section~\ref{sec:fromfactor}.

\section{Extracting $\eta$ and $\eta'$}

Since the configurations are well suited to study 
the effects of flavor symmetry breaking,
we start with the following SU(3) basis to describe $\eta$ and $\eta'$:
\begin{align}
 \eta_8 &= \frac{1}{\sqrt{6}}(u\bar{u} + d\bar{d} - 2s\bar{s}), 
 &
  \eta_1 &= \frac{1}{\sqrt{3}}(u\bar{u} + d\bar{d} + s\bar{s}).
\end{align}
The physical $\eta$ and $\eta'$ should be a mixing of 
the above octet state ($\eta_8$) and the singlet state ($\eta_1$):
\begin{align}
 \eta &=  \cos \theta\, \eta_8 - \sin \theta\, \eta_1 ,
 &
 \eta'&=  \sin \theta\, \eta_8 + \cos \theta\, \eta_1.
 \label{eq:eta-mixing}
\end{align}
We use a single angle parameterization and 
do not include a gluonic state in eq.~(\ref{eq:eta-mixing}),
since we expect the gluonic contributions to be small.
In principle we can resolve the contribution but is beyond
the current analysis.
The interpolating operator $\mathcal{O}_{\eta^{(')}}^\dagger$
that will be used
creates an overlap with the $\eta^{(')}$ state 
sufficient to obtain the target matrix elements.

\begin{figure}
 \centering
 \includegraphics[width=0.55\linewidth]{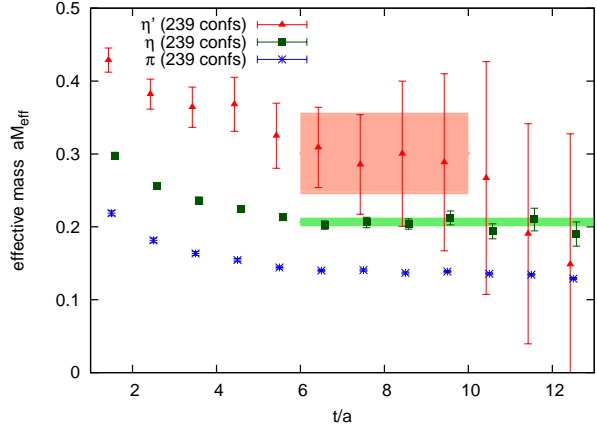}
 \caption{Effective mass plot for $\pi$, $\eta$, and $\eta'$
  for the $M_\pi
 = 348\, {\rm MeV}$ ensemble.}
 \label{fig:effmass}
\end{figure}

To extract the $\eta$ and $\eta'$ states from the lattice data, 
we need to diagonalize the following $2\times 2$ two point correlation matrix:
\begin{equation}
 C_2(t;\vec{k})=
 \begin{pmatrix}
  \langle \mathcal{O}_8(t;\vec{k})\mathcal{O}_8^\dagger(0) \rangle
  & \langle \mathcal{O}_8(t;\vec{k})\mathcal{O}_1^\dagger(0) \rangle
  \\
  \langle \mathcal{O}_1(t;\vec{k})\mathcal{O}_8^\dagger(0) \rangle
  & \langle \mathcal{O}_1(t;\vec{k})\mathcal{O}_1^\dagger(0) \rangle
 \end{pmatrix},
 \label{eq:2pt-2x2}
\end{equation}
where $\mathcal{O}_8$ and $\mathcal{O}_1$ are interpolating operators for 
$\eta_8$ and $\eta_1$, respectively, and $t$ is the time separation
between source and sink.  The operators at the sink are projected to
momentum $\vec{k}$. 
Here we use smeared interpolating operators 
(Wuppertal smearing \cite{Gusken:1989ad})
to reduce the effects from the excited states.
The diagonalized basis gives the operator $\mathcal{O}_{\eta^{(')}}$ 
for physical $\eta^{(')}$,
and the diagonalized two point functions give the spectra for $t/a \gg 1$:
\begin{equation}
 C_2^{{\rm diag},\ \eta}(t,\vec{k}) 
  = \frac{|Z_{\eta,\vec{k}}|^2}{2E_{\eta,\vec{k}}} \exp(-E_{\eta,\vec{k}} t),
 \qquad
 C_2^{{\rm diag},\ \eta'}(t,\vec{k}) 
  = \frac{|Z_{\eta'\vec{k}}|^2}{2E_{\eta',\vec{k}}} \exp(-E_{\eta',\vec{k}} t),
  \label{eq:2pt-diag}
\end{equation}
where 
$Z_{\eta^{(')},\vec{k}}
 = \langle \eta^{(')}(\vec{k})| \mathcal{O}_{\eta^{(')}}^\dagger(\vec{k})|0\rangle$ 
is an overlap factor between the $\eta^{(')}$ state with momentum $\vec{k}$
and a state created by the interpolating operator.
$E_{\eta^{(')},\vec{k}}$ is the energy of the state.
Setting $\vec{k}=\vec{0}$, we obtain the masses.
Note that each element in eq.~(\ref{eq:2pt-2x2}) contains both connected
and disconnected fermion loops.  For example, the $\eta_8 \to \eta_8$ two
point function contains the following fermion diagrams:
\begin{align*}
  \langle \mathcal{O}_8(t;\vec{k})\mathcal{O}_8^\dagger(0) \rangle
&= \textstyle \frac{1}{3}\Bigl[  \DiagTwoPtConn{l}{l} + 2 \DiagTwoPtConn{s}{s} \\
 & \textstyle -2 \bigl(\DiagTwoPtDisconn{l}{l}\bigr)
    -2 \bigl(\DiagTwoPtDisconn{s}{s}\bigr) 
  \textstyle +2 \bigl(\DiagTwoPtDisconn{l}{s}\bigr)
    +2 \bigl(\DiagTwoPtDisconn{s}{l}\bigr) \Bigr],
\end{align*}
where $l$ stands for light (i.e., $u$- and $d$- ) quarks.
In figure~\ref{fig:effmass}, we show the effective mass plot 
\begin{equation}
 a M_{\rm eff}\left(t+\frac{a}{2}\right)
  = \log\frac{C_2^{\rm diag}(t)}{C_2^{\rm diag}(t+a)}
\end{equation}
for $\eta$ and $\eta'$ together with that of $\pi$.
The figure demonstrates our ability to separate
the $\eta$ and $\eta'$ states.

The diagonalization presented 
so far gives a mixing of the interpolating operators 
$\mathcal{O}_1$ and $\mathcal{O}_8$, 
which depends on the smearing of our choice.
To obtain the \emph{physical} mixing, we replace the smeared operators 
at $t$ in eq.~(\ref{eq:2pt-2x2}) by local operators.
The smeared operators are still used at $t=0$ so we need to
use $Z_{\eta^{(')}}$ from smeared(source)-smeared(sink) two point functions 
to normalize the matrix elements.
Setting the momentum
$\vec{k}=\vec{0}$,  we can extract the following matrix elements:
\begin{align}
a_{\eta}^8= \langle 0 |\mathcal{O}^{\rm local}_8|\eta\rangle, &&
a_{\eta'}^8= \langle 0 |\mathcal{O}^{\rm local}_8|\eta'\rangle, &&
a_{\eta}^1= \langle 0 |\mathcal{O}^{\rm local}_1|\eta\rangle, &&
a_{\eta'}^1= \langle 0 |\mathcal{O}^{\rm local}_1|\eta'\rangle.
\end{align}
Modulo renormalization factors, 
they are proportional to the decay constants in the SU(3) flavor basis:
for example, $Z_8 a_{\eta}^8 = M_\eta f_{\eta}^8$, where $Z_8$ is the
renormalization factor and $f_{\eta}^8$ is the decay constant of $\eta$
through the octet.
We use the following ratio 
to obtain the mixing angle $\theta$,
\begin{equation}
 \tan^2 \theta 
  = \frac{\langle 0 |\mathcal{O}^{\rm local}_1|\eta\rangle
          \langle 0 |\mathcal{O}^{\rm local}_8|\eta'\rangle}
	 {\langle 0 |\mathcal{O}^{\rm local}_8|\eta\rangle
          \langle 0 |\mathcal{O}^{\rm local}_1|\eta'\rangle},
\end{equation}
for which the renormalization factors cancel.

\begin{figure}
 \centering
 \includegraphics[width=0.49\linewidth]{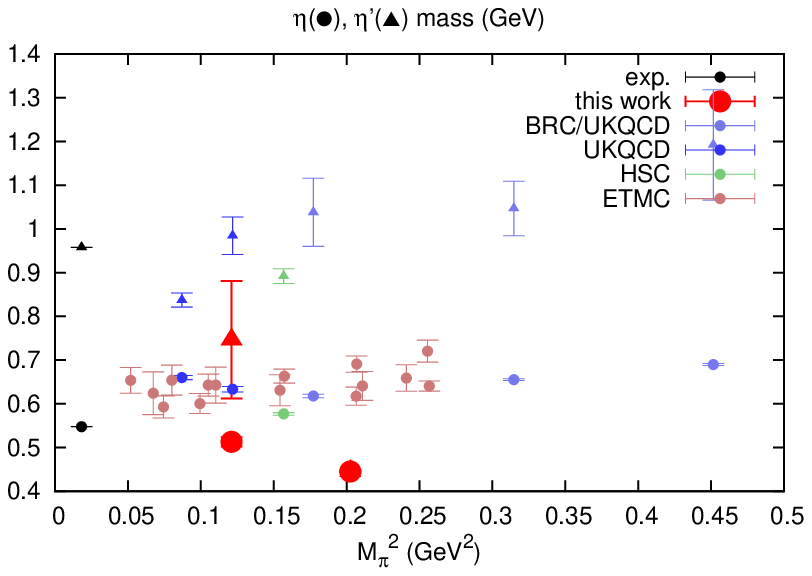}%
 \hfil%
 \includegraphics[width=0.49\linewidth]{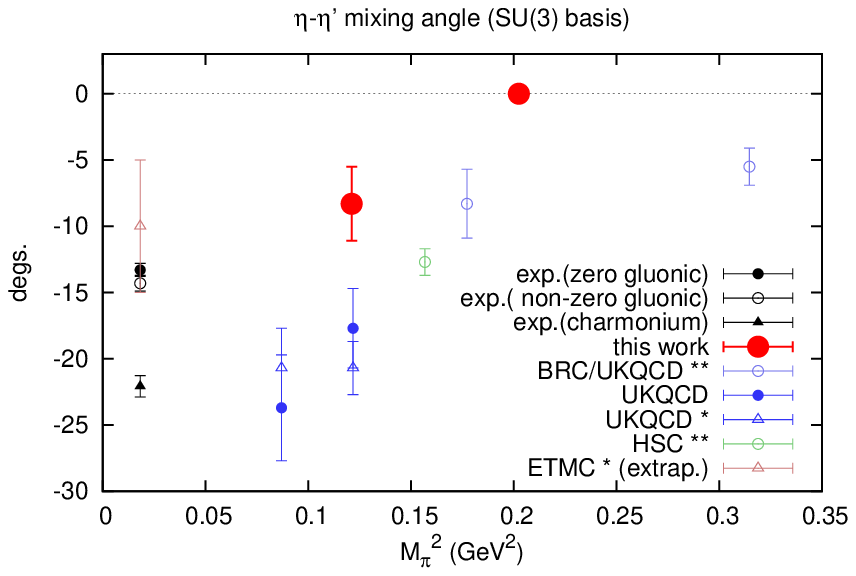}%
 \caption{Summary of lattice results for 
 the masses of $\eta$ and $\eta'$ (left panel)
 and the mixing angle (right panel).
 Included are result from
 the BRC/UKQCD \cite{Christ:2010dd}, 
 UKQCD \cite{Gregory:2011sg}.
 HSC \cite{Dudek:2011tt},
 and ETMC \cite{Ottnad:2012fv} collaborations. 
 For the mixing angle, two experimental values by KLOE
 \cite{Ambrosino:2009sc} ($\bullet$, $\circ$) 
 are from radiative decays 
$(\rho, \omega, \phi)\to  (\pi^0,\eta,\eta')\gamma$,
$\eta'\to\rho\gamma$ and$(\pi^0,\eta')\to \gamma\gamma$ ;
 they assume zero ($\bullet$) and non-zero ($\circ$) 
gluonic contributions to $\eta'$ state, respectively.
The other experimental value ($\blacktriangle$) is from BES 
\cite{Ablikim:2005je},  which uses charmonium decays only.
 Data with * use both local and fuzzed (smeared) operators and ** 
 fuzzed (smeared) operators only.
 }
 \label{fig:mass_and_mixing}
\end{figure}

Figure \ref{fig:mass_and_mixing} shows our preliminary values 
for the mass (left panel) and mixing angle (right panel) together with
the experimental values and other lattice results.
We find $M_\eta=513(11)\, {\rm MeV}$,
$M_{\eta'}=750(130)\, {\rm MeV}$, and $\theta=-8.3^\circ (2.8)$ 
at $M_\pi=348\, {\rm MeV}$.
Note that our result has no mixing 
by definition at the SU(3) flavor symmetric point: $\eta\equiv \eta_8$ and
$\eta' \equiv \eta_1$.
Currently, we have not yet included disconnected two point functions
for this ensemble, which means $M_\eta = M_{\eta'}$

\section{Decay Form Factors}
\label{sec:fromfactor}

Above we have constructed suitable interpolating operators 
for $\eta$ and $\eta'$,
so it is straightforward to obtain the matrix element we need
from the following three point function:
\begin{align}
 C_3^{D_s \to \eta^{(')}}(t)
 &= \langle \mathcal{O}_{\eta^{(')}}(t_f;\vec{k})
 S(t;\vec{q})\mathcal{O}_{D_s}^\dagger(t_i) \rangle
 \nonumber \\
 &\xrightarrow{t_f \gg t \gg t_i} 
 \frac{Z_{\eta^{(')},\vec{k}}\, Z_{D_s,\vec{p}}}
   {4 E_{\eta^{(')},\vec{k}}\, E_{D_s,\vec{p}}}
   \exp\left[-E_{D_s,\vec{p}}(t-t_i) - E_{\eta^{(')},\vec{k}}(t_f-t)\right] 
   \langle \eta^{(')})(\vec{k})|S(\vec{q})|D_s(\vec{p})\rangle.
 \label{eq:3pt-full}
\end{align}
Here, $\vec{p}=\vec{k} +\vec{q}$ and 
$\mathcal{O}_{D_s}$ is an interpolating operator for $D_s$.
We use a similar parameterization to eq.~(\ref{eq:2pt-diag})
for the two point function 
$\langle \mathcal{O}_{D_s}(t;\vec{p})\mathcal{O}_{D_s}^\dagger(0)\rangle$.
By combining eqs.~(\ref{eq:2pt-diag}), (\ref{eq:3pt-full})
and $\langle \mathcal{O}_{D_s}(t;\vec{p})\mathcal{O}_{D_s}^\dagger(0)\rangle$,
we obtain the matrix element
 $\langle \eta^{(')})(\vec{k})|S(\vec{q})|D_s(\vec{p})\rangle$.

\begin{figure}
 \centering
 \includegraphics[width=0.6\linewidth]{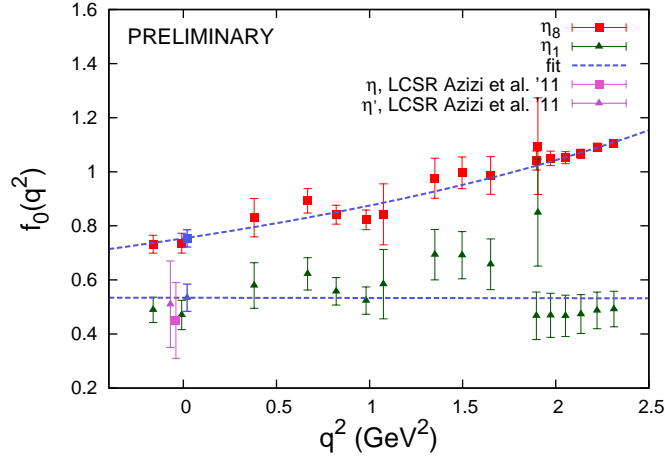}

 \caption{Preliminary results for the scalar form factor $f_0(q^2)$
 for the SU(3) flavor symmetric ensemble ($M_\pi=450\, {\rm MeV}$).
 Also included are results
 from light cone QCD sum rules (LCSR) \cite{Azizi:2010zj}.
}
 \label{fig:formfactor}
\end{figure}

In figure~\ref{fig:formfactor}, we plot our preliminary results for
$f_0(q^2)$ for the SU(3) flavor symmetric case.
A fit to the functional form 
$f_0(q^2)=\frac{f_0(0)}{1-bq^2}$ is also shown.
We find $f_0(0)=0.75(3)$
for $D_s \to \bar{l}\nu_l \eta$
and
$f_0(0)=0.52(5)$ for $D_s \to \bar{l}\nu_l \eta'$,
 at $M_\pi=450\ {\rm MeV}$.

To calculate the three point functions, 
we included all disconnected fermion loops.
As mentioned in the previous section, 
we have used the approximation
$M_\eta=M_{\eta'}$ to compute the two point functions,
which allows us to use different 
$t_{\rm sep}=t_f-t_i$ in eq.~(\ref{eq:3pt-full})
for the connected fermion loop contribution ($t_{\rm sep}=24a$) 
and the disconnected one ($t_{\rm sep}=8a$).
A larger $t_{\rm sep}$ suppresses possible pollution from excited
states, but leads to larger statistical errors, which is particularly
problematic for the disconnected diagrams.
Using different $t_{\rm sep}$ for the connected and disconnected three
point functions circumvents these problems in this case.
A full analysis with $M_\eta \neq M_{\eta'}$, which could
affect the $D_s \to \bar{l}\nu_l\eta_1$ form factor,
is currently under investigation. 
Due to flavor symmetry, the $D_s \to \bar{l}\nu_l\eta_8$ form factor 
is identical to $D\to \bar{l}\nu_l\pi$ and $D\to \bar{l}\nu_l K$
at the symmetric point ($M_\pi = 450\,{\rm MeV}$).

\begin{figure}
 \centering
 \includegraphics[width=0.6\linewidth]{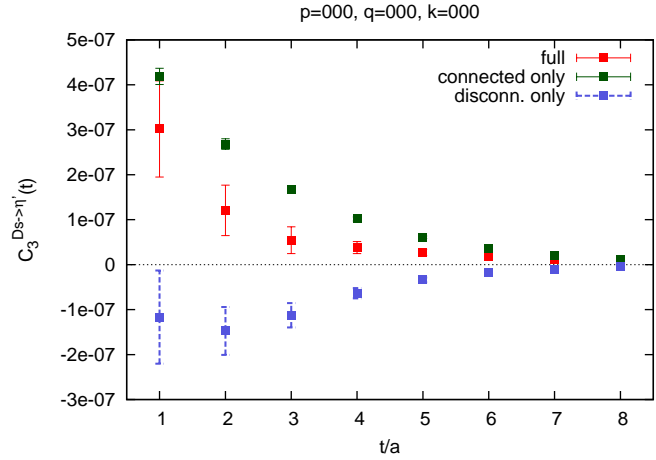}
 \caption{Connected and disconnected contributions to the three
 point function
$C_3^{D_s \to \eta'}(t)$ with
 $\vec{p}=\vec{q}=\vec{k}=\vec{0}$ and $M_\pi = 348\, {\rm MeV}$.
  The $D_s$ meson is created at $t/a=0$ and
 the $\eta'$ is annihilated at $t/a=8$.}
 \label{fig:conn_vs_disconn}
\end{figure}

Interestingly, we find that the contributions 
from the disconnected fermion loops are significant,
see figure~\ref{fig:conn_vs_disconn} where
both connected and disconnected contributions to 
$C_3^{D_s\to \eta'}(t)$ are plotted separately.
The figure also indicates
that the main source of the statistical 
error comes from the disconnected contribution.

The results presented here are still preliminary.  
In particular, the effects
of excited states of the $D_s$ and possibly $\eta$, $\eta'$
are still under investigation.
The effective mass plot (figure~\ref{fig:effmass}) suggests
that the excited state may contributes up to $t/a \lesssim 5$.
Since our fit region for the three point function
is at most $1\leq t/a \leq 7$ for the disconnected diagrams, 
there may be some pollution from excited states.
Note that the effects of an excited state (e.g. $\eta^*$) 
to the two point functions is suppressed 
by a square of the overlapping factor.
For example, including 
the leading order of the excited contributions eq.~(\ref{eq:2pt-diag})
becomes
\begin{equation}
 C_2^{{\rm diag},\eta}(t,\vec{k})
   = \frac{|Z_{\eta,\vec{k}}|^2}{2E_{\eta,\vec{k}}} \exp(-E_{\eta,\vec{k}} t)
     \left( 1+ 
      \frac{|Z_{\eta^*,\vec{k}}|^2}{|Z_{\eta,\vec{k}}|^2}
      \frac{E_{\eta,\vec{k}}}{E_{\eta^*,\vec{k}}}
      \exp\left[-(E_{\eta^*,\vec{k}} -E_{\eta,\vec{k}})t\right]
      +\cdots
      \right).
\end{equation}
The term $|Z_{\eta^*}/Z_{\eta}|^2$ appears as a correction.
On the other hand, the pollution to the three point function
is suppressed by only a linear factor:
\begin{align}
 C_3^{D_s \to \eta}(t)
 &=
 \frac{Z_{\eta,\vec{k}}\, Z_{D_s,\vec{p}}}
   {4 E_{\eta,\vec{k}}\, E_{D_s,\vec{p}}}
   \exp\left[-E_{D_s,\vec{p}}(t-t_i) - E_{\eta^{(')},\vec{k}}(t_f-t)\right] 
   \langle \eta(\vec{k})|S(\vec{q})|D_s(\vec{p})\rangle \nonumber\\
 & \times \Biggl[
    1+  \frac{Z_{\eta^*,\vec{k}}}{Z_{\eta,\vec{k}}}
      \frac{E_{\eta,\vec{k}}}{E_{\eta^*,\vec{k}}}
      \exp\left[-(E_{\eta^*,\vec{k}}-E_{\eta, \vec{k}})(t_f-t) \right]
      \frac{\langle \eta^*(\vec{k})|S(\vec{q})|D_s(\vec{p})\rangle}
           {\langle \eta(\vec{k})|S(\vec{q})|D_s(\vec{p})\rangle}
    \nonumber\\
 &\qquad\quad
   +
      \frac{Z_{D_s^*,\vec{p}}}{Z_{D_s,\vec{p}}}
      \frac{E_{D_s,\vec{p}}}{E_{D_s^*,\vec{p}}}
      \exp\left[-(E_{D_s^*,\vec{p}}-E_{D_s, \vec{p}})(t-t_i) \right]
      \frac{\langle \eta(\vec{k})|S(\vec{q})|D_s^*(\vec{p})\rangle}
           {\langle \eta(\vec{k})|S(\vec{q})|D_s(\vec{p})\rangle}
   + \cdots
\Biggr],
\end{align}
where the correction starts with the $Z_{\eta^*}/Z_{\eta}$ term.
It is important to remove these effects or
to estimate their magnitude, which may be one of the main sources
of the systematic error in our calculation.

\section{Conclusions}

We calculated the semi-leptonic decay form factors for 
$D_s \to \bar{l}\nu_l\eta$ and $D_s \to \bar{l}\nu_l \eta'$ using 
QCDSF $n_f=2+1$ lattice configurations.
This is the first lattice result of decay form factors
which includes the fermion disconnected loop
contributions.  
It turned out that the disconnected fermion loops give 
significant contributions to the form factor.  
The result is still preliminary but promising: 
the scalar form factor at zero momentum transfer $f_0(0)$ 
can be obtained with
$10$--$15$ \% precision.

We are planning to use larger lattices and lower quark masses.
Another target in addition to $f_0(q^2)$ is the other form factor 
$f_+(q^2)$.  We are also interested in calculating the 
$D_s\to \bar{l}\nu_l \phi$
decay form factor
that also contains a disconnected quark line contribution 
and the same calculation techniques are applicable.

\subsection*{Acknowledgements}

This work was supported by the DFG SFB/TR 55 and the EU ITN STRONGnet.
We used a modified version of CHROMA software suite \cite{Edwards:2004sx}.
The calculations were performed on the Athene HPC cluster and iDataCool
at the University of Regensburg.
It is my pleasure to thank my colleagues from the QCDSF collaboration
for providing the configurations.


\begin{thebibliography}{99}

\bibitem{DiDonato:2011kr}
  C.~Di Donato, G.~Ricciardi and I.~Bigi,
  Phys.\ Rev.\ D {\bf 85} (2012) 013016
  [arXiv:1105.3557 [hep-ph]].

\bibitem{Yelton:2009aa}
  J.~Yelton {\it et al.}  [CLEO Collaboration],
  Phys.\ Rev.\ D {\bf 80} (2009) 052007
  [arXiv:0903.0601 [hep-ex]].

\bibitem{Yelton:2010js}
  J.~Yelton {\it et al.}  [CLEO Collaboration],
  Phys.\ Rev.\ D {\bf 84} (2011) 032001
  [arXiv:1011.1195 [hep-ex]].

\bibitem{Na:2009au}
  H.~Na, C.~T.~H.~Davies, E.~Follana, P.~Lepage and J.~Shigemitsu,
  PoS LAT {\bf 2009} (2009) 247
  [arXiv:0910.3919 [hep-lat]].


\bibitem{Azizi:2010zj}
  K.~Azizi, R.~Khosravi and F.~Falahati,
  J.\ Phys.\ G {\bf 38} (2011) 095001
  [arXiv:1011.6046 [hep-ph]].


\bibitem{Witten:1979vv}
  E.~Witten,
  Nucl.\ Phys.\ B {\bf 156} (1979) 269.


\bibitem{Veneziano:1979ec}
  G.~Veneziano,
  Nucl.\ Phys.\ B {\bf 159} (1979) 213.


\bibitem{Bali:2011yx}
  G.~S.~Bali {\it et al.}  [QCDSF Collaboration],
  PoS LATTICE {\bf 2011} (2011) 283
  [arXiv:1111.4053 [hep-lat]].

\bibitem{Evans:2010tg}
  R.~Evans, G.~Bali and S.~Collins,
  Phys.\ Rev.\ D {\bf 82} (2010) 094501
  [arXiv:1008.3293 [hep-lat]].

\bibitem{Bali:2009hu}
  G.~S.~Bali, S.~Collins and A.~Sch\"afer,
  Comput.\ Phys.\ Commun.\  {\bf 181} (2010) 1570
  [arXiv:0910.3970 [hep-lat]].

\bibitem{Bietenholz:2010jr}
  W.~Bietenholz {\it et al.}[QCDSF collaboration],
  Phys.\ Lett.\ B {\bf 690} (2010) 436
  [arXiv:1003.1114 [hep-lat]].

\bibitem{Bietenholz:2011qq}
  W.~Bietenholz {\it et al.} [QCDSF collaboration],
  Phys.\ Rev.\ D {\bf 84} (2011) 054509
  [arXiv:1102.5300 [hep-lat]].

\bibitem{Cundy:2009yy}
  N.~Cundy {\it et al.},
  Phys.\ Rev.\ D {\bf 79} (2009) 094507
  [arXiv:0901.3302 [hep-lat]].

\bibitem{Gusken:1989ad}
  S.~G\"usken, U.~L\"ow, K.~H.~M\"utter, R.~Sommer, A.~Patel and K.~Schilling,
  Phys.\ Lett.\ B {\bf 227} (1989) 266.


\bibitem{Christ:2010dd}
  N.~H.~Christ {\it et al.},
  Phys.\ Rev.\ Lett.\  {\bf 105} (2010) 241601
  [arXiv:1002.2999 [hep-lat]].

\bibitem{Gregory:2011sg}
  E.~B.~Gregory {\it et al.}  [UKQCD Collaboration],
  Phys.\ Rev.\ D {\bf 86} (2012) 014504
  [arXiv:1112.4384 [hep-lat]].


\bibitem{Dudek:2011tt}
  J.~J.~Dudek, R.~G.~Edwards, B.~Joo, M.~J.~Peardon, D.~G.~Richards  and C.~E.~Thomas,
  Phys.\ Rev.\ D {\bf 83} (2011) 111502
  [arXiv:1102.4299 [hep-lat]].

\bibitem{Ottnad:2012fv}
  K.~Ottnad {\it et al.}  [ETM Collaboration],
  JHEP {\bf 1211} (2012) 048
  [arXiv:1206.6719 [hep-lat]].

\bibitem{Ambrosino:2009sc}
  F.~Ambrosino, A.~Antonelli, M.~Antonelli, F.~Archilli, P.~Beltrame, G.~Bencive
  JHEP {\bf 0907} (2009) 105
  [arXiv:0906.3819 [hep-ph]].

\bibitem{Ablikim:2005je}
  M.~Ablikim {\it et al.}  [BES Collaboration],
  Phys.\ Rev.\ D {\bf 73} (2006) 052008
  [hep-ex/0510066].

\bibitem{Edwards:2004sx}
  R.~G.~Edwards and 
B.~Jo\'o [SciDAC and LHPC and UKQCD Collaborations],
  Nucl.\ Phys.\ Proc.\ Suppl.\  {\bf 140} (2005) 832
  [hep-lat/0409003].

\end{thebibliography}
\end{document}